\documentclass[aps,prl,superscriptaddress,amsmath,amssymb,floatfix,twocolumn]{revtex4-1}
\usepackage{times}
\usepackage{graphicx,graphics,color}
\usepackage[colorlinks,citecolor=blue,linkcolor=red]{hyperref}
\usepackage{color}

\begin{document}

\title{Competing Electronic Configurations for PuTe and New Insight on Plutonium Metal}

\author{J. J. Joyce}
\email{jjoyce@lanl.gov}
\affiliation{Materials Science and Technology Division, Los Alamos National Laboratory, MS E574, Los Alamos, NM 87545, USA}

\author{ K. S. Graham}
\affiliation{Materials Science and Technology Division, Los Alamos National Laboratory, MS E574, Los Alamos, NM 87545, USA}

\author{J.-X. Zhu}
\email{jxzhu@lanl.gov}
\affiliation{Theoretical Division, Los Alamos National Laboratory,
Los Alamos, New Mexico 87545, USA}
\affiliation{Center for Integrated Nanotechnologies, Los Alamos National Laboratory,
Los Alamos, New Mexico 87545, USA}

\author{G. H. Lander}
\email{lander@ill.fr}
\affiliation{Directorate for Nuclear Safety and Security, European Commission, Joint Research Centre, Postfach 2340, D-76125 Karlsruhe, Germany}

\author{H. Choi} 
\affiliation{Theoretical Division, Los Alamos National Laboratory,
Los Alamos, New Mexico 87545, USA}

\author{T. Durakiewicz}
\affiliation{Institute of Physics, University of Maria Curie-Sklodowska, 20-031 Lublin, Poland}

\author{J. M. Wills}
\affiliation{Theoretical Division, Los Alamos National Laboratory,
Los Alamos, New Mexico 87545, USA}

\author{P. H. Tobash}
\affiliation{Materials Science and Technology Division, Los Alamos National Laboratory, MS E574, Los Alamos, NM 87545, USA}

\author{ E. D. Bauer}
\affiliation{Materials Physics and Applications Division, Los Alamos National Laboratory, MS K764,  Los Alamos, NM 87545, USA}

\author{J. N. Mitchell}
\affiliation{Materials Science and Technology Division, Los Alamos National Laboratory, MS E574, Los Alamos, NM 87545, USA}

\begin{abstract}
The electronic structure of plutonium metal and its compounds pose a grand challenge for a fundamental understanding of the Pu-5$f$ electron character. For 30 years the plutonium chalcogenides have been especially challenging, and multiple theoretical scenarios have been proposed to explain their unusual behavior. We present extensive high-resolution photoemission data on a single crystal of PuTe, which has also been proposed as a topological insulator. The new experimental results on this mixed-valent material provide a constraint to the theoretical modeling and new dynamical mean-field theory calculations agree with the experimental results. Comparisons with Pu metal provide new insight in understanding its complex electronic structure. 

\end{abstract}

\maketitle

Plutonium metal and its compounds exhibit a wide range of spectacular properties, including the highest superconducting transition temperature ($T_c \sim 18$ K in PuCoGa$_5$) of any heavy fermion material~\cite{JLSarrao:2002}.
The source of the rich electronic structure in Pu materials is the 5$f$ electrons and the ability of these electrons to assume multiple valence configurations. Not only is there compelling evidence of mixed valence in Pu metal and compounds, but there is also a localized/itinerant boundary for the character of the 5$f$ electrons~\cite{TGouder:2000,TDurakiewicz:2004,PSoderlind:2004,JCLashley:2005,JJJoyce:2006,JHShim:2007,JXZhu:2007,vanderLaan:2010,CHBooth:2012,MJanoschek:2015}. PuTe sits at one endpoint for the electronic structure range of Pu compounds with more 5$f^6$ character than any other solid-state Pu compound. As a result, there have been many efforts to provide a theory to explain the unusual properties of this material~\cite{PWachter:1991,PMOppeneer:2000,LPetit:2002,LVPurovskii:2005,AOShorikov:2005,ASvane:2006,AShick:2007,AShick:2009,FBultmark:2009,MTSuzuki:2009,CHYee:2010,MMatsumoto:2011,XZhang:2012}. Recently,  a strong temperature dependence of the electronic structure~\cite{CHYee:2010} and a topological insulating state~\cite{XZhang:2012} of PuTe have been predicted.  The latter proposal becomes extremely interesting for the following reason: Most of the topological insulators are spin-orbit coupling driven but with weak electronic correlations, which can be described accurately by first-principles electronic structure method within density functional theory; whereas the $f$-electron materials have not only strong spin-orbit coupling but also strong electronic correlation effects, which can give rise to new phenomena. So far, SmB$_6$ is one of the most compelling examples of strongly correlated topological insulators~\cite{MDzero:2010,MDzero:2016} and the possible topological insulator in 5$f$-electron based compounds with even stronger spin-orbit coupling has attracted increased attention recently~\cite{XDeng:2013,HChoi:2018}.  

The purpose of this Letter is to understand the balance between 5$f^5$ and 5$f^6$~\cite{PWachter:1991,PMOppeneer:2000,LPetit:2002,LVPurovskii:2005,AOShorikov:2005,ASvane:2006,AShick:2007,AShick:2009,FBultmark:2009} and test the above mentioned theoretical predictions~\cite{CHYee:2010,XZhang:2012}.  The electronic structure of PuTe (NaCl fcc crystal structure) is dominated by the valence 5$f$ electrons. The experimental data shown in this paper use high-resolution ($\Delta E=12$ meV) angular-resolved photoemission on a cleaved single crystal, measured with the sample temperature varied from 8-300 K~\cite{KSGraham:2013}.

\begin{figure}[h]
\centering
\includegraphics[width=0.8\linewidth,height=0.75\textheight,clip]{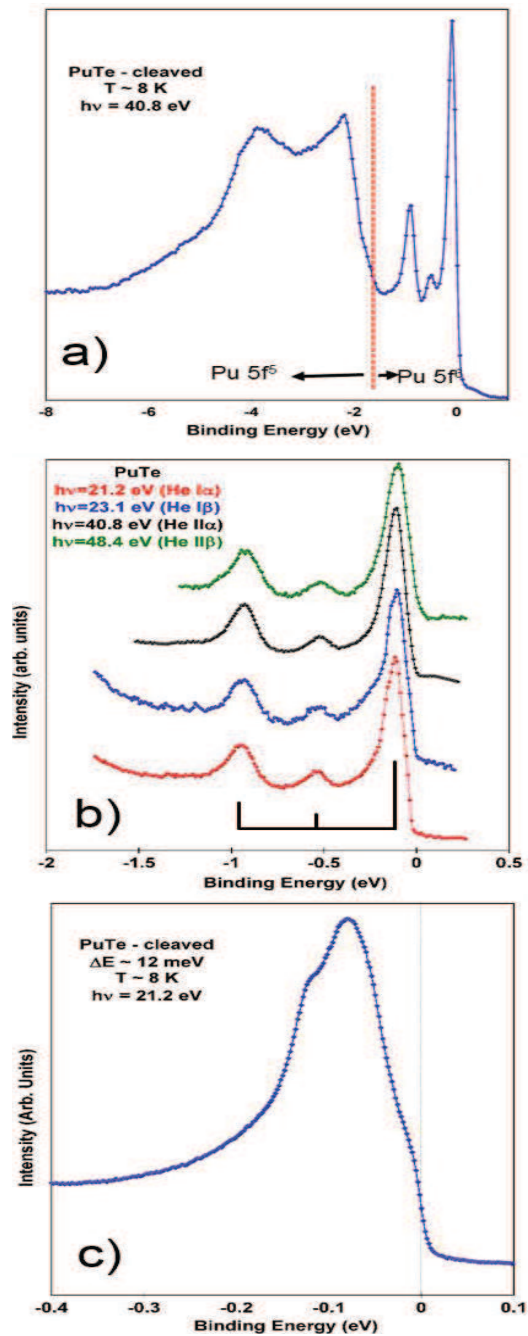}
\caption{(Color online) PES data for PuTe showing a) the full valence band with 5$f^5$ and 5$f^6$ configurations; b) the `3 peak structure' in the first 1 eV energy interval invariant in photon energy; c) the first peak near $E_F$ with a binding energy of $\sim$ 70 meV and indication of a small second component. Energy resolution for panels a) and b) is $\Delta E=120$ meV, and $30\sim 50$ meV, respectively.
}
\label{FIG:PES-PuTe}
\end{figure}

In Fig.~\ref{FIG:PES-PuTe} we show the valence band of PuTe over 3 different binding-energy intervals and 4 different incident-photon energies. The energy intervals cover the full range of the valence-band region and the fine details of the Fermi energy, whereas the incident photon energies cover a broad cross-section (Pu 5$f$, Pu 6$d$, Te 5$p$ are the principle orbitals) that switch between a dominance of the conduction band at low photon energy, and a dominance by the 5$f$ states at higher photon energy. Comparing cross-sections at 21.2 eV and 40.8 eV we see the ratio Pu 5$f$/6$d$ ~ 0.4 at 21.2 eV and this ratio jumps to 13.6 by a photon energy of 40.8 eV (Pu 5$f$/Te 5$p$ ratios $\sim$ 0.3 and 19)~\cite{JJYeh:1985}.  

Figure~\ref{FIG:PES-PuTe}(a) shows that the signal from the 5f electrons of Pu indicate two discrete configurations, 5$f^5$ and 5$f^6$, as first observed in the isostructural and isoelectronic compound PuSe by Gouder {\em et al.}~\cite{TGouder:2000}. Figure~\ref{FIG:PES-PuTe}b shows the agreement of the 3-peak structure with atomic multiplet theory~\cite{FGerken:1983}. This arises from a localized 5$f^6$ initial state transitioning into a 5$f^5$ final state in the photoemission process. The three peaks are shown in Fig.~\ref{FIG:PES-PuTe}(b) as a function of incident-photon energy from 21 eV through 48 eV; their relative intensity and energy positions are independent of incident photon energy. This invariance, when the orbital cross-sections vary by factors of 40 at these photon energies, suggests~\cite{JJYeh:1985,FGerken:1983} that the entire 3-peak manifold is of pure 5$f$ character with little hybridization. Further indication of the atomic multiplet nature of these peaks is given by comparison to the atomic calculations~\cite{FGerken:1983}, which are represented by the three black vertical bars below the data in Fig.~\ref{FIG:PES-PuTe}(b). Spanning an energy interval of almost 1 eV, our PES data and the calculations agree within 8 meV, or better than one percent. Additionally, the comparison between calculation and PES data agree in the relative intensity of these three peaks to within six percent.

The second region in Fig.~\ref{FIG:PES-PuTe}(a) with binding energy below $-2$ eV is much broader in width and represents the Pu 5$f^5$ configuration and is similar in lineshape and energy position to the localized Pu 5$f$ levels in the magnetic material PuSb~\cite{TDurakiewicz:2004}. The lower energy (i.e. further from $E_F$) part of the curve also contains a contribution around $-4$ eV from the Te 5$p$ states~\cite{AShick:2009}. The separation in energy ($\sim$ 1 eV) of the two 5$f$ configurations is reminiscent of the separation observed in 4$f$ configurations in rare-earth mixed-valent materials~\cite{MCampagna:1974,MGrioni:1985,JJJoyce:1996}. We assume these fluctuations are dynamic in nature, as is the case in similar rare-earth materials~\cite{MCampagna:1974,MGrioni:1985,JJJoyce:1996}. If they were static the lattice parameter would not be close to that for trivalent PuSb~\cite{PMOppeneer:2000}.

Figure~\ref{FIG:PES-PuTe}(c) shows the high resolution PES data for PuTe with just the first peak of the `3 peak structure'. With an energy resolution of 12 meV, the binding energy of this first peak is determined to be 70 meV below $E_F$ with a full-width at half maximum of $\sim$ 140 meV. The Fermi level is clearly visible in this frame and it is also apparent that the 5$f$ peak is cut by the Fermi energy at the measurement temperature of 8 K. The high-resolution data shows evidence of a small second component in this first peak. Line shape analysis (See Supplemental Material (SM) Fig. S1~\cite{SI:2018}) shows this second component is $\sim$ 8\% of the total  intensity of the strongest first peak.

Overall, Fig.~\ref{FIG:PES-PuTe} shows an electronic structure for PuTe that is mixed valent with components of 5$f^5$ and 5$f^6$ configurations. The 5$f$ occupancy ($n_f$) is defined as the average number of 5$f$ electrons, which is a number between 5 and 6 with a reasonable estimate of about 5.5. The 5$f^6$ component is a localized 3 peak structure well described by atomic multiplet theory and the 5$f^5$ component is similar to the 5$f^5$ identified in PuSb. Both contributions are removed from $E_F$. There is a model where the mixed-valent nature of the fluctuations explains why the lattice parameter~\cite{PMOppeneer:2000} mirrors that of a localized 5$f^5$ system (it is only 1\% smaller than that of PuSb), the absence of magnetism with a temperature independent susceptibility~\cite{PWachter:1991}, and why the neutron form factor is different from that for 5$f^5$\cite{GHLander:1987}, and why there is a low conductance~\cite{JMFournier:1990}  and a small Sommerfeld coefficient~\cite{GRStewart:1991}. The model is also consistent with the drop in resistivity under pressure~\cite{VIChas:2001} at 11 GPa, when the 5$f^6$ state is presumably suppressed, and there is a sign of a magnetic transition at 15 K, consistent with magnetism in a localized 5$f^5$ system.

\begin{figure}[t]
\centering
\includegraphics[width=1.0\linewidth,clip]{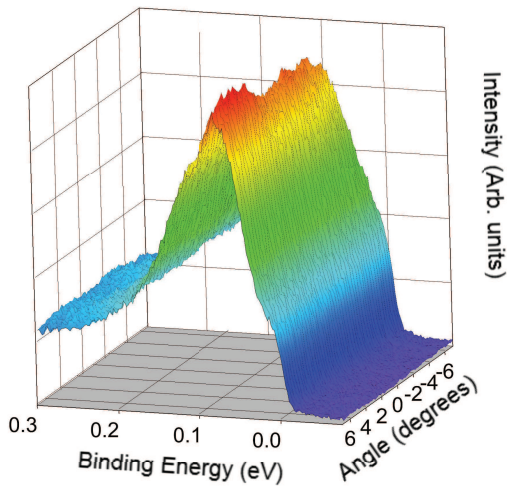}
\caption{(Color online) ARPES data for PuTe at 21.2 eV showing no dispersion or surface state. Energy resolution $\Delta E  = 12$ meV.  ARPES at other incident energies is shown in the SM Fig. S2~\cite{SI:2018}. 
}
\label{FIG:ARPES-PuTe}
\end{figure}

A recent publication~\cite{XZhang:2012} predicting topological insulator states in PuTe and AmN was the motivation for our angle-resolved photoemission (ARPES) studies of PuTe. In Fig.~\ref{FIG:ARPES-PuTe} we show ARPES data taken at a photon energy of 21.2 eV. There are two important points arising from this ARPES data. First, there is no evidence of a surface state in the APRES data that could potentially be a topologically protected state. Such a surface state is expected to lead to a small but measurable local increase of density of states near the Fermi level in photoemission spectra, as shown for a 4$f$ mixed-valent system SmB$_6$~\cite{MNeupane:2013}. Neither angle-integrated, nor angle-resolved spectra shown here contain such features for PuTe. The lack of a surface-like state in photoemission precludes the existence of a topological insulator state. Second, there is no crystal momentum dependence of states observed in the ARPES data. The data for an incident photon energy of 21.2 eV is shown in Fig.~\ref{FIG:ARPES-PuTe}, while additional ARPES data at 23.1 and 40.8 eV are shown in Fig. S2 of the SM~\cite{SI:2018}. In all, we investigated four different photon energies and five different angle ranges to cover a sizable portion of reciprocal space in our investigation. We observed no evidence of a topological state and no dispersion in the electronic structure of this first peak. The basis for the prediction in Ref.~\onlinecite{XZhang:2012} is that the materials PuTe and AmN would have an actinide 5$f^6$ configuration. Our present experiments clearly establish that PuTe has a mixed-valent ground state, some of which can be associated with the 5$f^6$ configuration, but clearly this is not sufficient to induce the topological insulator surface states.

Whereas the ARPES data for PuTe provide no evidence of a topological insulator state, it does provide compelling evidence for the localization of the Pu 5$f^6$ configuration representing the electronic structure near the Fermi energy. Unlike Pu materials such as PuCoGa$_5$ and PuSb$_2$, which have shown dispersion in the ARPES data for valence states near $E_F$~\cite{KSGraham:2013,JJJoyce:2010}, PuTe is without crystal momentum dependence in the peak nearest E$_F$.

\begin{figure}[t]
\centering
\includegraphics[width=0.8\linewidth,clip]{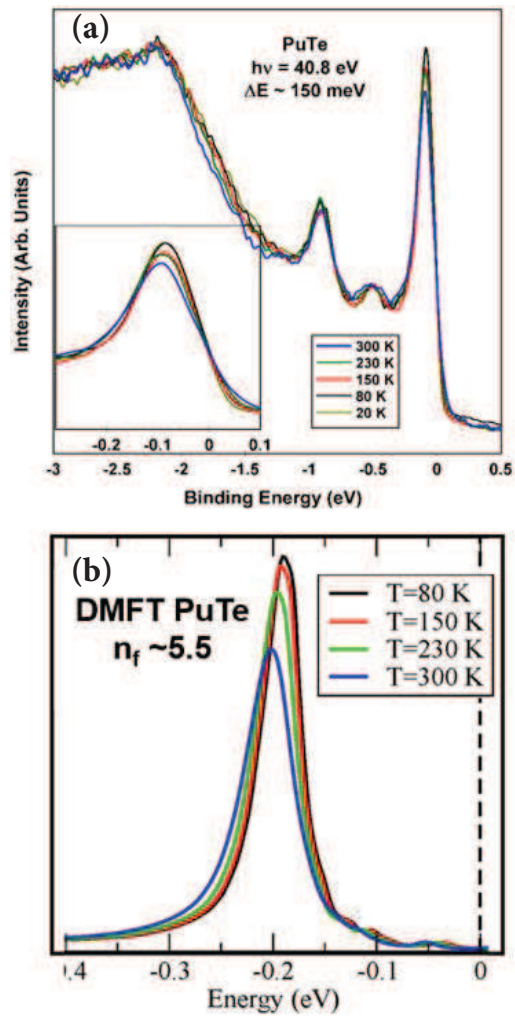}
\caption{(Color online) (a) Temperature dependent PES for PuTe and (b) LDA+DMFT predictions on the temperature dependence of spectral density for PuTe. The inset to panel (a) 
shows a zoomed-in view of the first peak below the Fermi energy.
}
\label{FIG:T-Dept}
\end{figure}

\begin{figure}[t]
\centering
\includegraphics[width=0.8\linewidth,keepaspectratio,clip]{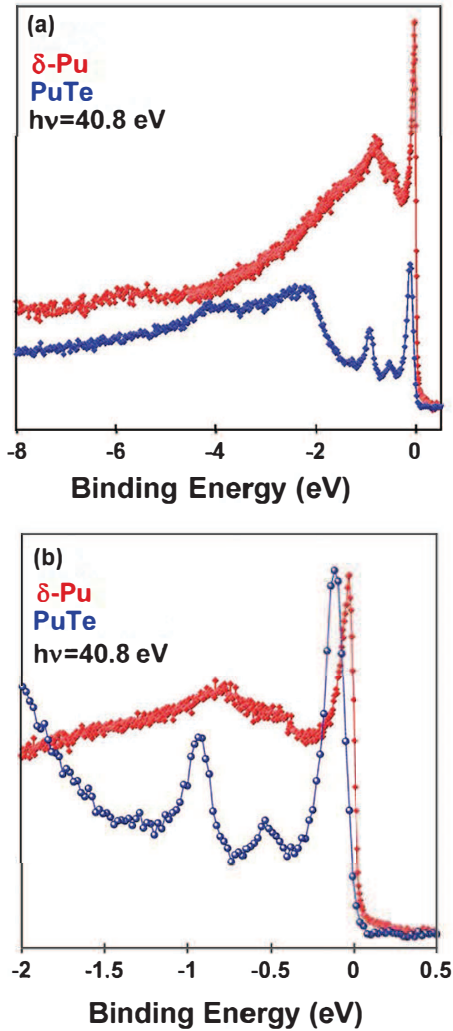}
\caption{(Color online) High-resolution ($\Delta E \sim 60$ meV) photoemission of PuTe compared to that of $\delta$-Pu (stabilized with $\sim$ 2\% of Ga) with (a) full valence band and (b) features within 2 eV of the Fermi level. 
In (a) the scaling has been done at 4 eV, which corresponds to the lower part of the valence band. In (b) the scaling is at the highest intensity of the first peak near $E_F$. 
}
\label{FIG:PES-Pu}
\end{figure}

We have already mentioned many of the theoretical efforts on Pu-chalcogenides~\cite{PWachter:1991,PMOppeneer:2000,LPetit:2002,LVPurovskii:2005,AOShorikov:2005,ASvane:2006,AShick:2007,AShick:2009,FBultmark:2009,MTSuzuki:2009,CHYee:2010,MMatsumoto:2011,XZhang:2012}, and one of the more recent work using the combined local density approximation with dynamical mean-field theory (LDA+DMFT) method~\cite{CHYee:2010} had the additional feature that it predicted a strong temperature dependence of the electronic structure. To test this prediction we have measured the temperature-dependence of the spectra at various incidence photon energies from PuTe, and these are shown (at 40.8 eV) in Fig.~\ref{FIG:T-Dept}(a). No temperature dependence, apart from Fermi-Dirac statistics,  of the intrinsic electronic structure at any photon energy was observed. Specifically, fitting of the PES data with a peak centered 70 meV below $E_F$ and a linewidth of $\sim$ 140 meV is consistent with the temperature dependence of our PuTe data convoluted with the appropriate Fermi function for the temperature range 20 to 300 K, without any significant components exhibiting temperature dependence, which might come from, for example, many-body interactions. We also note that our observation of temperature-independent spectra is robust against the incident photon energy as shown in Fig. S3 of the SM~\cite{SI:2018}.

Since the publication of Ref.~\onlinecite{CHYee:2010},  there has been a convergence in the LDA+DMFT modeling on both $\alpha$ and $\delta$ phases of Pu metal~\cite{JXZhu:2013}. 
In Fig.~\ref{FIG:T-Dept}(b) we show LDA+DMFT calculations resulting from the more recent collaborative efforts on Pu LDA+DMFT study~\cite{MJanoschek:2015}.  The $n_f$ value for the PuTe in Fig.~\ref{FIG:T-Dept}(b) is $\sim5.5$, which is significantly larger than the value of 5.2 reported in Ref.~\onlinecite{CHYee:2010}. The spectral density with this value of $n_f$ agrees well with the experiment on its temperature dependence. A series of calculations were performed with $n_f$ values ranging from 5.0 to 5.5 with systematics showing a stronger temperature dependence as the $n_f$ value goes down (See SM Fig. S4~\cite{SI:2018}). 

Finally, we turn to the significance of our results in terms of understanding the electronic structure of Pu metal, about which there has been a long-standing controversy~\cite{JJJoyce:2006,JHShim:2007,JXZhu:2007,vanderLaan:2010,CHBooth:2012,MJanoschek:2015}. We compare the spectral response of PuTe, as discussed in this paper, with that of $\delta$-Pu, on which the most work has been done to understand the electronic structure.
The overall signal across the valence band is shown in Fig.~\ref{FIG:PES-Pu}(a), and the regions within 2 eV below $E_F$ is shown in Fig.~\ref{FIG:PES-Pu}(b). It is immediately clear that the peak in $\delta$-Pu is much sharper than that in PuTe, and it is located much closer to the Fermi level in $\delta$-Pu metal. This confirms that in $\delta$-Pu the 5$f$ electrons near the Fermi energy are mostly itinerant, and have their spectral weight close to $E_F$.
 This is consistent with a metal, and the long-standing understanding that the 5$f$ electrons are centered at or very near $E_F$ in $\delta$-Pu. On the other hand, it is also clear from this comparison that there is a remnant of the 5$f^6$ multiplet feature in $\delta$-Pu, but that it does not constitute nearly as much as in PuTe, and makes up no more than 20\% of the signal in $\delta$-Pu. Using the comparison with PuTe, we have a measure of the 5$f^6$ occupation in $\delta$-Pu as
 as $n_f = 5.2(1)$.

In conclusion, new high-resolution photoemission data for PuTe along with DMFT calculations provide insight into the electron structure of this end-point Pu material, resolving a long-standing controversy over the electronic structure~\cite{PWachter:1991,PMOppeneer:2000,LPetit:2002,LVPurovskii:2005,AOShorikov:2005,ASvane:2006,AShick:2007,AShick:2009,FBultmark:2009,MTSuzuki:2009,CHYee:2010,MMatsumoto:2011,XZhang:2012}.  The material is mixed valent with a 5$f$ contribution of close to 5.5. There is no evidence for either the proposed topological state~\cite{XZhang:2012}, or for the predicted strong temperature-dependence~\cite{CHYee:2010} of the electronic structure. There is an excellent quantitative agreement with the multiplet structure arising from a localized 5$f^6$ initial state decaying into a 5$f^5$ final state in the photoemission process. Crucially, the resolution of the electronic structure of PuTe in this work has allowed important comparisons with the well-known metallic $\delta$-Pu metal, leading to comprehensive restrictions on the amount of 5$f^6$ occupation that can be proposed for the metal.

\acknowledgements{
This work was
carried out under the auspices of the U.S. Department of Energy
(DOE) National Nuclear Security Administration under Contract No. 89233218CNA000001.
The photoemission research  were supported by the U.S. DOE Basic Energy Sciences (BES) through Materials Science and Engineering Division. The theoretical work was supported by  LANL LDRD  DR Program, and in part supported by Center for Integrated Nanotechnologies, a DOE BES user facility, in partnership with LANL Institutional Computing Program for computational resource. TD acknowledges the support of the NSF IR/D program. We gratefully acknowledge the PuTe crystals provided by the European Commission, Joint Research Centre, Karlsruhe, Germany. 
}


\end{document}


\title{Supplemental Material for Competing Electronic Configurations for PuTe and New Light on Plutonium Metal}

\author{J. J. Joyce}
\email{jjoyce@lanl.gov}
\affiliation{Materials Science and Technology Division, Los Alamos National Laboratory, MS E574, Los Alamos, NM 87545, USA}

\author{ K. S. Graham}
\affiliation{Materials Science and Technology Division, Los Alamos National Laboratory, MS E574, Los Alamos, NM 87545, USA}

\author{J.-X. Zhu}
\email{jxzhu@lanl.gov}
\affiliation{Theoretical Division, Los Alamos National Laboratory,
Los Alamos, New Mexico 87545, USA}
\affiliation{Center for Integrated Nanotechnologies, Los Alamos National Laboratory,
Los Alamos, New Mexico 87545, USA}

\author{G. H. Lander}
\email{lander@ill.fr}
\affiliation{Directorate for Nuclear Safety and Security, European Commission, Joint Research Centre, Postfach 2340, D-76125 Karlsruhe, Germany}

\author{H. Choi} 
\affiliation{Theoretical Division, Los Alamos National Laboratory,
Los Alamos, New Mexico 87545, USA}

\author{T. Durakiewicz}
\affiliation{Institute of Physics, University of Maria Curie-Sklodowska, 20-031 Lublin, Poland}

\author{J. M. Wills}
\affiliation{Theoretical Division, Los Alamos National Laboratory,
Los Alamos, New Mexico 87545, USA}

\author{P. H. Tobash}
\affiliation{Materials Science and Technology Division, Los Alamos National Laboratory, MS E574, Los Alamos, NM 87545, USA}

\author{ E. D. Bauer}
\affiliation{Materials Physics and Applications Division, Los Alamos National Laboratory, MS K764,  Los Alamos, NM 87545, USA}

\author{J. N. Mitchell}
\affiliation{Materials Science and Technology Division, Los Alamos National Laboratory, MS E574, Los Alamos, NM 87545, USA}

\maketitle


%



In this Supplemental Material (SM) document, we show details for several of the measurements and calculations presented in the main text. In Fig.~\ref{Sfig:PES-fitting} we show high resolution (12 meV energy resolution) ARPES data with lineshape analysis to add additional detail to the information presented with Fig. 1 in the main text. In Fig.~\ref{Sfig:ARPES-photon} we show ARPES data at three different photon energies, which expand and reinforce the lack of dispersion observed in the ARPES data for PuTe. Figure~\ref{Sfig:PES-TempDep}  adds an additional photon energy onto the temperature dependence of the PES data as presented in Fig. 3 of the main text. Likewise, Fig.~\ref{Sfig:DMFT-TempDep} adds additional $n_f$ values for the spectral density to complement that presented in Fig. 3b of the main text, which were calculated within the the combined local density approximation with dynamical mean-field theory (LDA+DMFT) method. The extended ranges of photon energies and $n_f$ values presented here reinforce the choice of photon energies and $n_f$ values selected for presentation in the main text as the best values, and demonstrate the range of experimental and theoretical values explored in order to arrive at the conclusions reached in the main text. Finally, Fig.~\ref{Sfig:DMFT-focc} shows the LDA+DMFT calculations with a wider energy range, which captures the full valence band region with the different $n_f$ values. This full valence band calculation may be compared against the full valence band data presented in Fig. 1a of the main text. The additional information presented in this SM documents demonstrates the larger parameter-space that  was explored both experimentally and theoretically to arrive at the stated conclusions in the main text.

\section{Photoemission spectroscopy on $\mathbf{PuTe}$}
\begin{figure}[th]
\centering\includegraphics[
width=1.0\linewidth,clip]{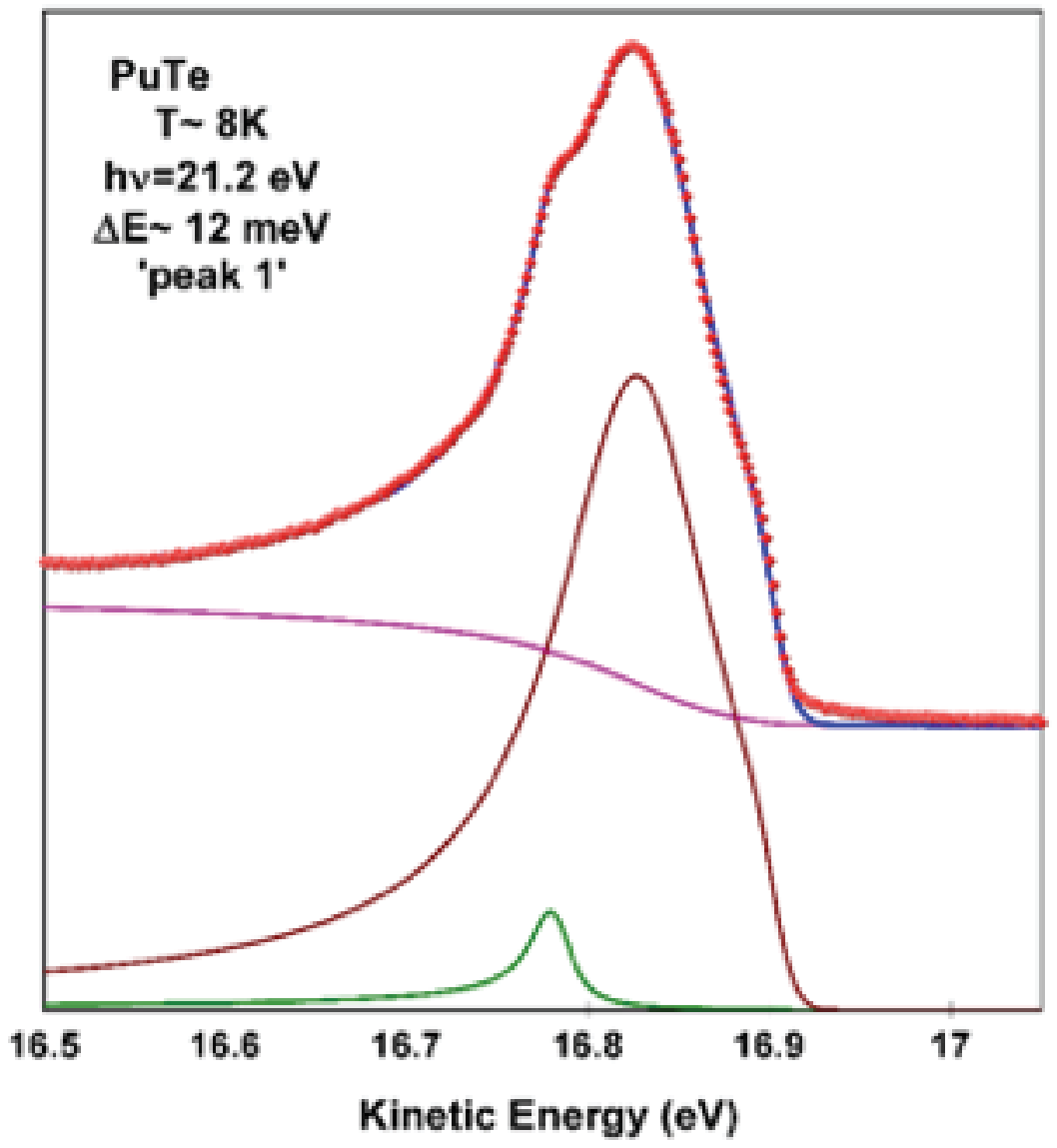}
\caption{(Color online) 
Fitting of ‘Peak1’ at 12 meV resolution to the measured PES. This shows how good the fitting is and how small the second component is even though the raw data has a visual effect of a major second component. The role of the Fermi function on the major component leads to the perception (incorrectly) that the second component (deeper binding energy) is a significant fraction of the area. In fact, it is only about 7\% of the integrated area.
}
\label{Sfig:PES-fitting}
\end{figure}

The electronic structure of PuTe is divisible into two main regions, which are representative of Pu 5$f$ levels consisting of a 5$f^5$ and a 5$f^6$ configuration. These two regions were delineated in Fig. 1a of the main text with a red line at a binding energy just below 2 eV. The 5$f^6$ configuration was further detailed with the ‘3-peak’ structure in Fig. 3b. With increased energy resolution, we showed that even this first peak of the ‘3-peak’ structure had a fine structure within this first component. In order to detail this fine structure, we performed lineshape analysis on the first peak observed in the high-resolution ARPES data. In Fig.~ \ref{Sfig:PES-fitting}, the red dots are the data, the black line is the sum of the fitted peaks and the three components that make up the black fitted line are the Shirley background function and two peaks. The two peaks are Gaussian-broadened Doniach-Sunjic lineshapes. The Gaussian broadening represents the 12 meV energy resolution for the measurement. The Doniach-Sunjic lineshape is a lineshape based on the Lorentzian lineshape which is the fundamental lineshape for an oscillator with  accommodations made for a low-energy asymmetric tail characteristic of a metallic state~\cite{JJJoyce:1989}. Several variations of the fitting were run but all variations of the fitting with a low $\chi$-squared value showed spectral weight in the range of 7-8\%  for the second (minor) component and the asymmetry of the lineshape was low indicating the core-hole was not in the presence of a strong metallic background giving rise to low energy electron-hole pairs which skew the low energy tail of a metallic energy level.

\begin{figure}[th]
\centering\includegraphics[
width=1.0\linewidth,clip]{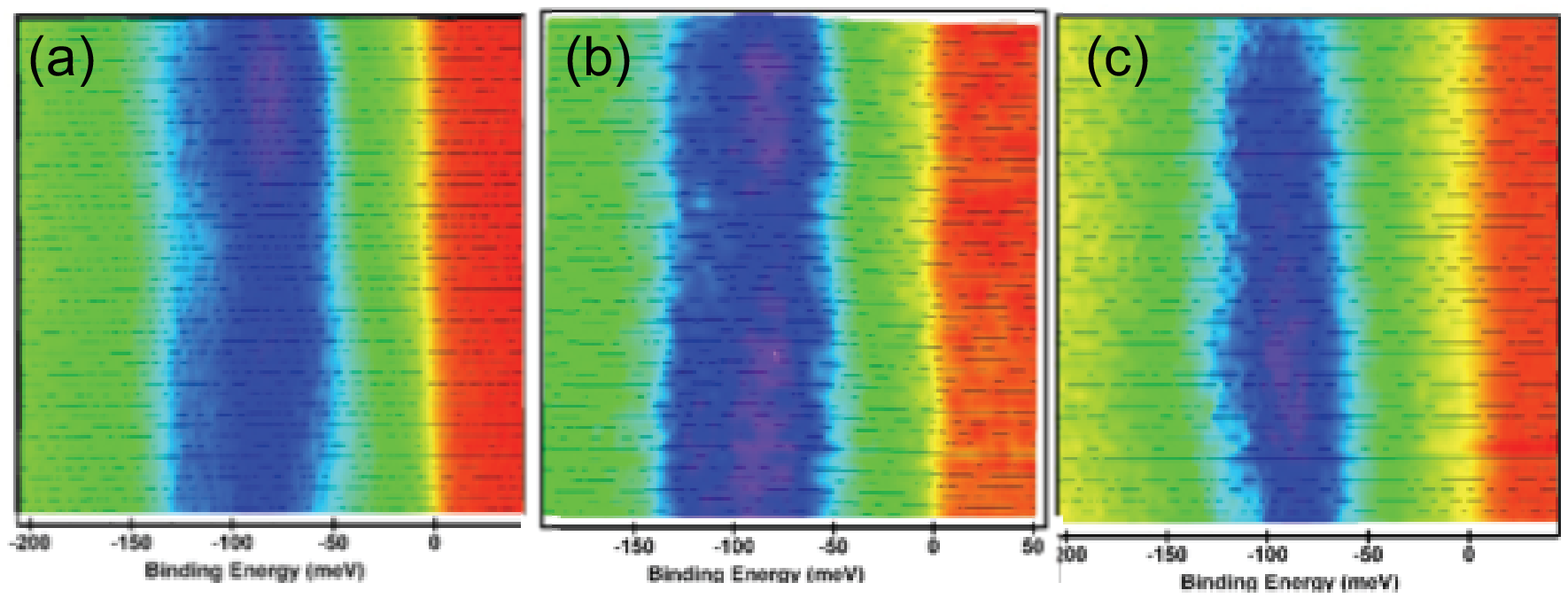}
\caption{(Color online) 
ARPES on PuTe at $T$= 8 K at 3 photon energies 21.2 eV (a), 23.1 eV (b) and 40.8 eV (c). The energy resolution $\Delta E \sim 12 \;\text{meV}$. It shows a greater coverage in $\bf{k}$-space than the figure in the main body of the paper. The total lack of dispersion at any energy makes a compelling argument for localized Pu-5$f$ electrons.
}
\label{Sfig:ARPES-photon}
\end{figure}

In Fig.~\ref{Sfig:ARPES-photon} we show $\mathbf{k}$-independence over three photon energies which probe different regions of $\mathbf{k}$-space and different cross-sectional variations for 5$f$ versus 6$d$ orbitals. The variation in photon energy is estimated to move the $k_z$ component of the ARPES data through roughly half of $k_z$  reciprocal space (the PuTe inner potential is indeterminate but by comparison to UTe, where the lattice constant is nearly identical and the inner potential was determined by sweeping the photon energy at a synchrotron over a wide photon energy interval). The three photon energies shown in Fig.~\ref{Sfig:ARPES-photon} are 21.2 eV, 23.1 eV and 40.8 eV. In all three frames there is no detectable dispersion in electronic structure which covers the first peak of the ‘3-peak’ Pu 5$f^6$ feature. The total lack of any dispersion is fully consistent with a localized 5$f^6$ state, which again is consistent with the excellent agreement between the ARPES data and the atomic structure calculations done in Ref.~\onlinecite{FGerken:1983}. The excellent agreement between the PES data and the fitted peaks leaves little room for substantive variations in the fitted lineshapes and relative ratios of integrated intensities. The origin of the 7-8\% fitted component has not been unambiguously identified at this point.

\begin{figure}[th]
\centering\includegraphics[
width=1.0\linewidth,clip]{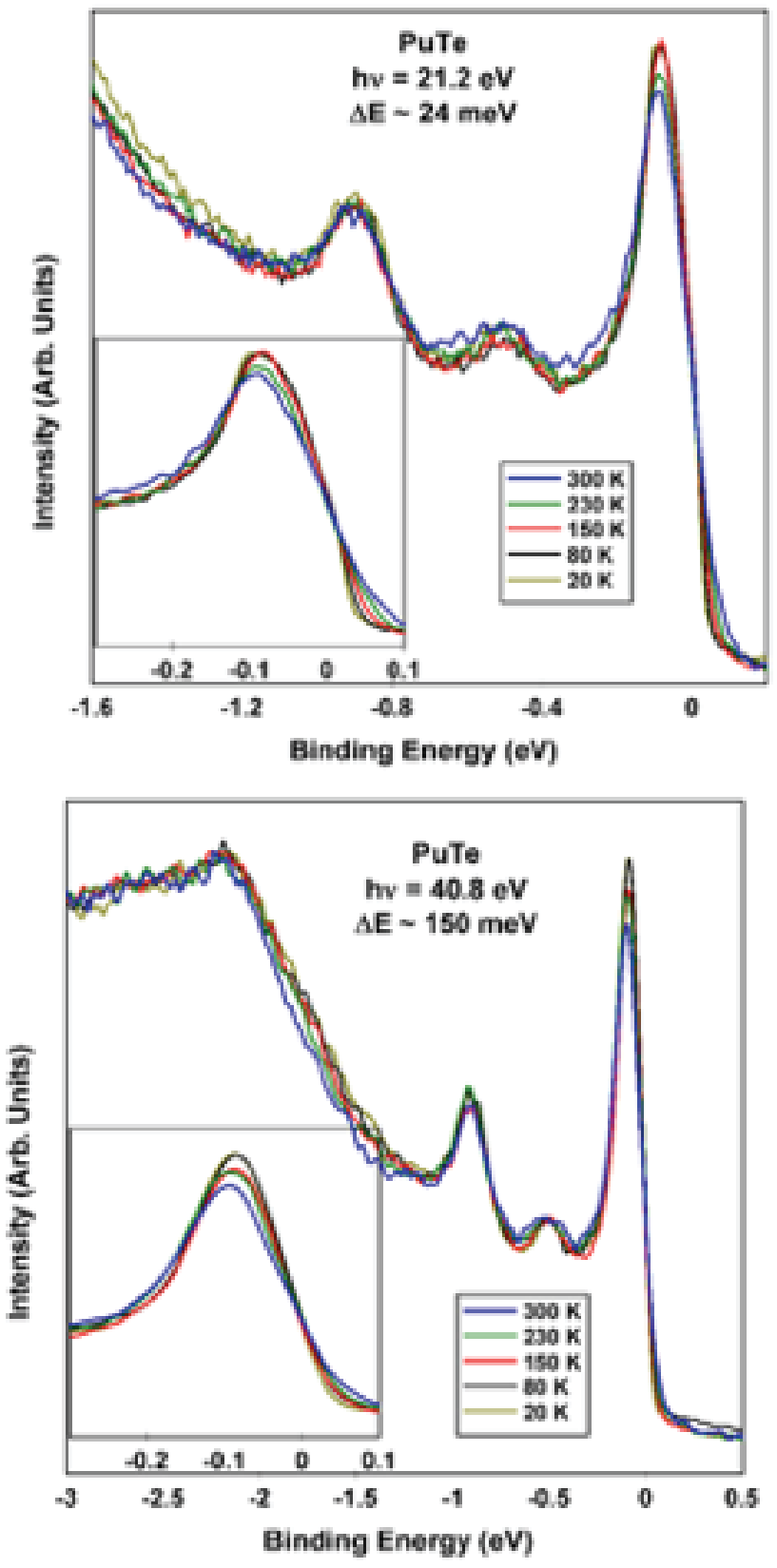}
\caption{(Color online) 
Temperature dependence of the ‘3-peak’ structure at two different energies 21.2 eV (a) and 40.8 eV (b). It shows the result is robust again the incident photo energy. It also shows that it is the same for different orbital cross-sections. Note that 21.2 eV is Pu-6$d$ sensitive while 40.8 eV is Pu-5$f$ dominant.
}
\label{Sfig:PES-TempDep}
\end{figure}

In Fig.~\ref{Sfig:PES-TempDep} we show temperature-dependent PES data at two photon energies to supplement the temperature-dependent data shown in Fig. 3a of the main text. The two photon energies show the same temperature dependence even though the Pu 5$f$ and 6$d$ orbital cross section ratios change by a factor of 40 with the two energies~\cite{JJYeh:1985}. At a photon energy of 21.2 eV the Pu 6$d$ states will have larger cross-section than the Pu 5$f$ levels  but by 40.8 eV the Pu 5$f$ cross-section dwarfs the other states and dominates the spectral intensity. The fact that the temperature dependence is the same at both photon energies speaks to the reproducibility of the temperature dependence as well as the orbital purity of the ‘3-peak’ structure, which dominates the electronic structure of PuTe in the first one eV from the Fermi energy. The insets for each photon energy in Fig.~\ref{Sfig:PES-TempDep} show details of the temperature dependence of the first peak about the Fermi energy. The shape and symmetry of this temperature dependence is fully consistent with Fermi-Dirac statistics for a narrow peak near (but not at) the Fermi energy undergoing temperature variations.

\section{LDA+DMFT Spectral Density of $\mathbf{PuTe}$}
The LDA+DMFT calculations  were performed using EDMFTF package~\cite{KHaule:2010} in connection with the full-potential linearized augmented plane-wave (FP-LAPW) based Wien2k code~\cite{PBlaha:2001}. One-crossing approximation was employed to solve the quantum impurity problem self-consistently within the DMFT.
We used  $U=4.5\;\rm{eV}$ for the Hartree component of the screened Coulomb interaction, which is consistent with previous work on elemental Pu~\cite{SYSavrasov:2001,JHShim:2007,JXZhu:2007,CHYee:2010,JXZhu:2013}. The remaining Slater integrals ($F^2$, $F^4$, and $F^6$) were calculated using Cowan's atomic structure code~\cite{RDCowan:1981}
 and reduced by 30\% to account for screening, which leads to the Hunds-rule exchange $J=0.512\;\rm{eV}$. We take the double-counting energy to be $E_{\rm{DC}}=U(n_{f}^{0}-1/2) - J(n_{f}^{0}-1)/2$ with $n_{f}^{0}$ as a varying parameter, which controls the final Pu-5$f$ occupancy.
Throughout the work,  all calculations were performed
at the experimentally determined lattice constants~\cite{OLKruger:1967}, $RK_{\text{max}}=8$, and a 12 $\times$ 12 $\times$ 12 $\mathbf{k}$-mesh in the Brillouin zone.

\begin{figure}[t]
\centering\includegraphics[
width=1.0\linewidth,clip]{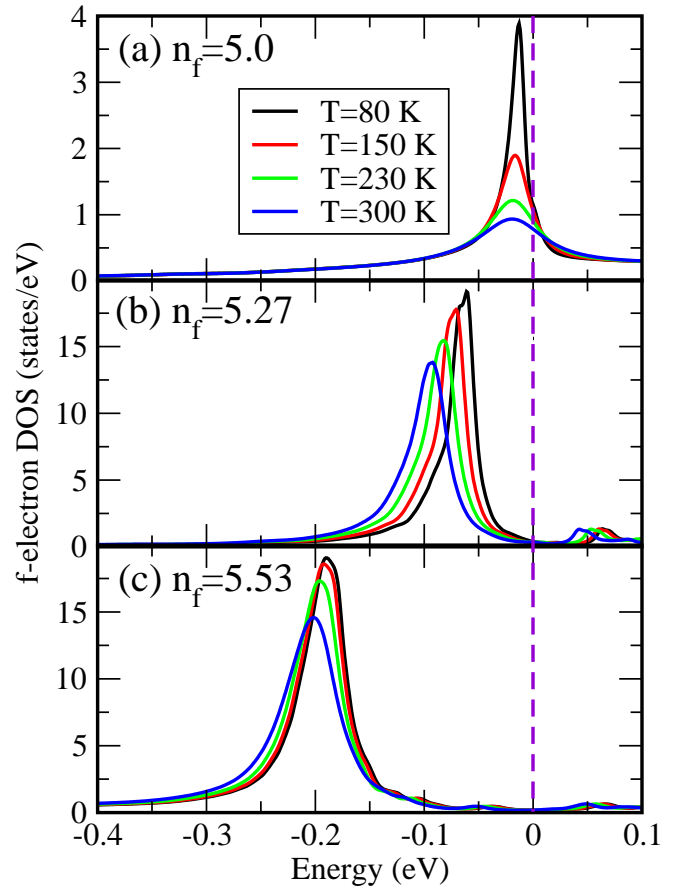}
\caption{(Color online) 
Temperature dependence of the $f$-electron density of states for varying $f$-electron occupation. It shows that a suppression of the quasiparticle peak at a fixed energy location (near the Fermi energy) for $n_f=5.0$, a quasiparticle peak (about 0.05 meV below the Fermi energy) shifting with varying temperature for $n_f=5.27$, as well as an almost temperature-independent quasiparticle peak (located about 0.2 eV below the Fermi energy) for $n_f=5.53$.
}
\label{Sfig:DMFT-TempDep}
\end{figure}

\begin{figure}[th]
\centering\includegraphics[
width=1.0\linewidth,clip]{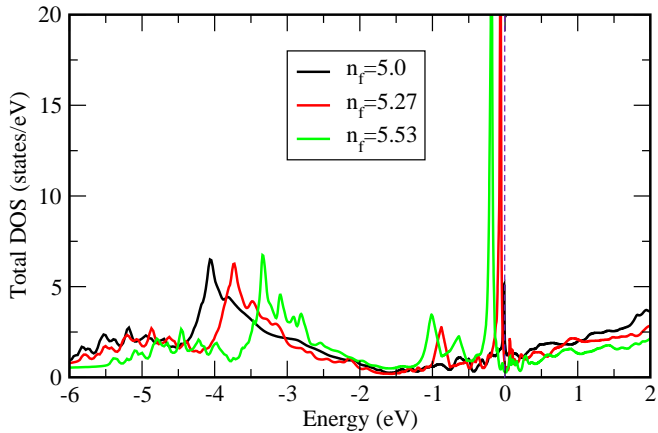}
\caption{(Color online) 
$f$-electron occupation dependence of the total density of states (DOS) at $T=80$ K. It shows a clear multiplet feature in the DOS  for $n_f=5.27$ and $5.53$. In particular, the three-peak structure within the energy range [-1.5,0] eV for $n_f=5.53$ gives the best agreement with the PES measurement. In addition, with the increasing Pu-5$f$ occupation, the Te-5$p$ states are shifted toward the direction of the Fermi energy, making their location also agreeing better with the experimental data when $n_f=5.53$. The strong agreement between theory with $n_f=5.53$ and experiment gives a compelling evidence that the PuTe is a highly mixed valent compound. 
}
\label{Sfig:DMFT-focc}
\end{figure}

In Fig.~\ref{Sfig:DMFT-TempDep} we show LDA+DMFT calculations for three values of $n_f$ ranging from 5.0 to 5.53. This range of values supplements the calculation shown in  Fig. 3b of the main text, where we presented the calculation which agrees best with the experimental results. With the extended range of $n_f$ values and calculations, we see very distinct differences in the temperature dependence of the features centered about the Fermi energy. First, the temperature dependence of the $n_f=5.0$ calculation has a huge predicted temperature variation. This is entirely inconsistent with the experimental data, which shows only a Fermi-Dirac temperature dependence for the electronic structure nearest the Fermi energy. It was the large temperature dependence predicted in Ref.~\onlinecite{CHYee:2010} that provided some of the impetus to carry out the experiments on PuTe. Interestingly, the LDA+DMFT calculation here for $n_f=5.0$ looks a great deal like the temperature dependence predicted for $n_f=5.2$~\cite{CHYee:2010}. Likewise the LDA+DMFT calculation for $n_f=5.27$ is also inconsistent with the experimental data as it predicts a shift in binding energy for the peak position as well as a stronger temperature dependence than is realized in the experimental data. The LDA+DMFT calculation for $n_f=5.53$ gives the result, which bears the strongest resemblance to the experimental data set -- It has the smallest temperature dependence and it shows not significant energy shifts in the primary feature with temperature.

In Fig.~\ref{Sfig:DMFT-focc} we show the LDA+DMFT calculations  over the full valence band region  which further delineate the calculations between  $n_f$ values of 5.0, 5.27 and 5.53 which support the model of the largest $n_f$ value (5.53) agreeing with the experimental data. In addition to the differences as a function of temperature and $n_f$ value noted in the discussion of Fig.~\ref{Sfig:DMFT-TempDep}, we note that in Fig.~\ref{Sfig:DMFT-focc}, there is a  distinct difference in the binding energy of the feature associated with the  5$f^5$ component in the PES/ARPES which is centered  a little over 2 eV below the Fermi energy in  Fig. 1a of the main text. It is noted that there are actually two features below 2 eV in  the experimental data,  these features have been identified by Shick {\em et al.}  as Pu 5$f$ character nearer the Fermi energy and ligand $p$-states (Te, Se, S) as indicated by systematics for the family of rocksalt Pu chalcogenides (PuTe, PuSe and PuS)~\cite{AShick:2009}. In the context of this identification that the Pu 5$f$ is nearer the Fermi energy in this manifold and the ligand $p$-states are deeper in binding energy, it would seem that the LDA+DMFT calculation with the  5$f^5$ states closest to 2 eV in binding energy would be most consistent with the experimental data (Fig. 1a of the main text). Once again, this comparison favors the larger $n_f$ value in the LDA+DMFT calculation for PuTe over the $n_f$ values closer to 5.0.

In summary, these five supplemental figures further support  the conclusions of our paper, which are that the electronic structure of PuTe has an $n_f$ value which is close to 5.5, that the  5$f^6$ manifold which dominates the electronic structure in the first one eV near the Fermi energy is localized and well described by final state multiplets and finally, the experimental and computational results for PuTe support the largest $n_f$ value yet observed in Pu  solid state samples with the greatest fraction of Pu 5$f^6$ electron character.